\documentclass[oneside,12pt]{article}

\usepackage{graphicx}

\usepackage{graphics}
\usepackage{hyperref}
\usepackage{amsmath}
\usepackage{graphicx}
\usepackage{amsfonts}
\usepackage{amssymb}

\def\i{{\it i}}
\def\p#1{\epsilon_#1}

\def\br{\mathop{\bf r}}
\def\brb{\mathop{\bf r_2}}
\def\bra{\mathop{\bf r_1}}
\def\bs{\mathop{\bf s}}
\def\bsa{\mathop{\bf s_1}}
\def\bsb{\mathop{\bf s_2}}
\def\bv{\mathop{\bf v}}

\def\p{\mathop{\partial}}
\def\half{{1\over 2}}

\def\br{\mathop{\bf r}}
\def\bv{\mathop{\bf v}}

\begin{document}
\title{ Models of breakup: a final state interaction problem.{\\ \small  In memory of Mahir Hussein.}}
\author{Angela Bonaccorso$^{1}$\footnote{\emph{Email:} bonac@df.unipi.it} ~and~ David M. Brink$^{2}$\footnote{\emph{Deceased}} \\ 
\small $^1$Istituto Nazionale di Fisica Nucleare, Sezione di Pisa, \\ \small Largo Bruno Pontecorvo 3, 56124 Pisa, Italy. \\ 
\small  $^2$Rudolf Peierls Centre of Theoretical Physics, University of Oxford, \\ \small 1 Keble Road, Oxford OX1 3NP, U.K.   }
%
%
\maketitle
\begin{abstract} {In this paper we discuss the evolution of breakup models from fully quantum mechanical, such as the Ichimura-Austern-Vincent model to semiclassical, to eikonal approximations  following   the insight on the mechanism first proposed by Hussein and McVoy (HM) for the presently called stripping term. In particular  we  concentrate on, and stress that,  the  correct implementation of a quantum mechanical model of breakup  requires the use of energy dependent interactions and the energy averaging procedure is a key point to understand the difference among models.  On the other hand  using fixed energy  potentials is one of the steps towards the high energy eikonal limit first proposed by HM. However the  intermediate semiclassical transfer to the continuum model  (STC) of Bonaccorso and Brink does use an energy dependent nucleon-target optical potential, while fixing the core-target interaction at the incident energy. The relationship between these methods is clarified.}\end{abstract}

PACS{24.10.Eq, 25.70.Mn, 25.45.-z}
\onecolumn

%
\section{Introduction}





\vspace{1em}

In the last forty years nuclear reaction theory  largely benefitted from the activity of Mahir Hussein (MH). Mahir had been a student of  A. Kerman and thus he came from one of the world-class  top-schools in the subject. In particular MH dedicated a large part of his research activity to the developments of direct reaction theory. Mahir had a gift for physics insight of complicated problems and one of his most notable achievements was to create   a tradition of semiclassical nuclear reaction models and a group of experts on it in Brasil.

Among direct reactions, breakup emerged as one of the most important  following the seminal papers of a number of authors.   They are mostly based on extensions of the Distorted Wave Born Approximation (DWBA) theory to
the case of an unbound final state.
The problem with DWBA is that it has a complicated formalism
and it is difficult to disentangle the various reaction mechanisms
which are believed to contribute to the inclusive spectra.
 The work of Huby et
al.,\cite{9),10)},  is similar  to our Semiclassical Transfer to the Continuum (STC)\cite{bb,bb1,67,initst,ppnp} since in both theories
the final state for the unbound nucleon is represented by a scattering
state with an appropriate normalization. The problem of normalization is of fundamental importance if one is interested in the absolute value of the breakup cross section and not only in the shape of the spectra. In fact since the advent of exotic nuclei,   measurements and calculations of absolute breakup cross sections \cite{PPNPall,Tostevin_Gade} have concentrated on the extraction of spectroscopic factors \cite{Macfarlane}. Other approaches
by Tamura and Udagawa\cite{6)}, Mermaz \cite{7),8)}
are based on statistical compound nucleus theories. They require
quite lengthly numerical calculations and do not determine the
absolute normalization. McVoy and Nemes \cite{13}, proposed a
simple model based on the Plane Wave Born Approximation to
calculate both transfer to continuum and break-up. However they
obtained only very qualitative results.

 Later Ichimura \cite{MI}, revised the existing approaches to what has been called 
 inclusive break-up, which is the sum of elastic and inelastic breakup (called also diffraction and stripping) corresponding to an  inclusive reaction in which only the core is detected  when a given projectile fragments on a target. These approaches are due to Udagawa and Tamura\cite{6)},
Austern and Vincent\cite{AV}, Kasano and Ichimura \cite{KI},   J. Pampus et al. \cite{PAMPUS1978141}, G. Baur et al. \cite{BAUR1984333}, and finally to Hussein and McVoy \cite{Hussein:1985}. These works mainly
differ in the model wave function used to describe the final state as we shall see in the following.

 In fact there are several ways in which a nucleon in a bound state in the projectile
 can make a transition to a continuum final state.
In all cases the nucleon ends up in an unbound state. The approaches differ
in their treatment of the final state interactions (FSI) and of the corresponding continuum nucleon wave function. One possibility is
transfer to a continuum state of the target. This means that final state
interactions of the nucleon with the target are included. Transfer to resonances
states of the target can be described by this approach. A second possibility
is to include FSI with the projectile-core and a third is to
neglect FSI  altogether. 
It is clear that these processes can not all be independent.
They are discussed in  detail in a formal way based on the time dependent G-matrix approach in Ref. \cite{bb1} and in  Appendix C of this work. 

Then, in the originally  complicated scenario, emerged the paper by Hussein and McVoy  \cite{Hussein:1985} in which a simple semiclassical reduction of the inclusive breakup formula was proposed in terms of a WKB approximation for the projectile and core distorted waves. Furthermore  averaging the energy dependence of the interactions and wave functions on the whole kinematically allowed range and thus taking the relative projectile-target, core-target, and nucleon target momenta and potentials at the value of the incident energy  per nucleon an eikonal formula was deduced.  Note that the inclusive breakup models  contain two terms, one for the elastic breakup and another one for the inelastic breakup. However the eikonal reduction for the EBU term was not attempted by HM. Thus a formal derivation of the eikonal formalism for EBU from one of the fully QM methods  is still missing.

 On the other hand Fujita and  H\"ufner \cite{FH}  and  H\"ufner and Nemes \cite{HN} started directly from an eikonal formalism to describe breakup at relativistic energies.  Later on, in  the '90s, somewhat independently Yabana et al.\cite{YABANA} and Hencken et al. \cite{Hencken:1996} introduced  eikonal  models to  study  halo breakup in reactions induced by radioactive nuclei. They  obtained in an almost straightforward way the  stripping term and  derived also the diffraction term but  for the integrated cross section they needed to introduce a further hypothesis besides the initial weak binding namely that the projectile should have no bound excited states. For diffraction they derived a formula    for the total  cross section and the intrinsic momentum distribution.  It should be noticed that HM did not use the weakly bound projectile hypothesis to deduce their model and obtained as we said before only the stripping term. Later on an eikonal formalism for core perpendicular and parallel momentum distributions was obtained by Bertulani et al. \cite{bertulani04}. Carstoiu et al.\cite{PhysRevC.70.054602} used a sudden approximation method to obtain both parallel and transverse distributions and absolute cross sections.

One of the  fully quantum mechanical QM models of breakup, introduced by Ichimura-Austen-Vincent (IAV)\cite{IAV85},   was originally complicated to understand from the formal point of view and also it was impossible to apply   in a numerical calculation unless the projectile was small and the energy low, because computers at that time were not able to handle the large number of partial waves necessary for heavy-ion reactions at high or even intermediate energies. Recently this model has been revised by Jin Lei and Moro \cite{Jin15,Jin18b,Jin17,Jin15b,Jin18} and other groups \cite{Pot17,Carlson2016,Pot15} who have  been able to implement it numerically and made several applications, still at low energy and for deuteron or $\alpha$ breakup.

Thus, it is now well understood that considering the FSI of the breakup nucleon  with the target means that the nucleon can re-scatter on the target elastically and inelastically. The two possibilities correspond obviously to the elastic and inelastic cross sections of a free-nucleon-target interaction as described by the optical model  for example. This suggests that physically  the most  natural choice of the nucleon-target final wave function is a continuum  wave function determined by the optical model. Because the nucleon in the continuum can have an energy from zero to a maximum allowed by kinematical conditions, it is clear that the  final wave function must be energy dependent. Thus  the correct implementation of the model requires an energy dependent   optical potential as stressed by IAV \cite{IAV85}. However an energy dependent potential while  applied in the past to low energy reactions and/or small projectile-target combinations \cite{Jin15,Carlson2016,Pot15},  had not been implemented until recently  \cite{jinme} in any  of the QM model mentioned above in the case of heavy ion reactions at high energy (\textgreater 50A.MeV). On the other hand the  standard  implementation of the STC method is with an energy dependent potential because it was introduced for heavy-ion reactions at medium to high incident energies where the energy spectra are quite broad \cite{Highensp,Tina}. Recently results from the IAV and STC models have been compared \cite{jinme} calculating neutron and proton breakup from $^{14}$O and $^{16}$C on a $^9$Be target \cite{Flavigny:2012}. This is the first example of an application of the IAV model implemented with an energy dependent potential, for heavy ions at intermediate energies.

In  Sec. 2 of this paper we shall begin by  giving IAV and  HM formulae and linking them to the STC formalism and we shall show explicitly how the choice of an appropriate final state wave function and potential come about. Energy averaging will be also discussed and  in Sec. 3 we will give some examples of energy spectra  and integrated cross sections according to the STC and eikonal models and clarify the limit of applicability of the each of them. Finally in Sec. 4 our conclusions will be drawn.

\section{IAV, HM and STC models.}




\subsection{IAV model}
In this section, we briefly summarize the model of Ichimura, Austern, and Vincent (IAV), whose original derivation can be found in Refs.~\cite{IAV85,AUSTERN1987125} and has been also recently revisited by several authors~\cite{Jin15,Jin18b,Jin15b,Jin18,Pot17,Carlson2016,Pot15,jinme}. We outline here the main results of this model and refer the reader to these references for further details on their derivations. The discussion and notation follows that of the recent publication \cite{jinme} in which the IAV and STC method were compared.

We write the reaction under study in the form,
\begin{equation}
P(=C+n)+T \rightarrow C+B^{*}, 
\end{equation}
where the projectile $P$, composed of $C$ and $n$, collides
with a target $T$, emitting $C$ fragments and any other fragments. Thus, $B^{*}$ denotes any final state of the $n+T$ system.

\begin{figure}[tb]
\begin{center}
 {\centering \resizebox*{0.56\columnwidth}{!}{\includegraphics{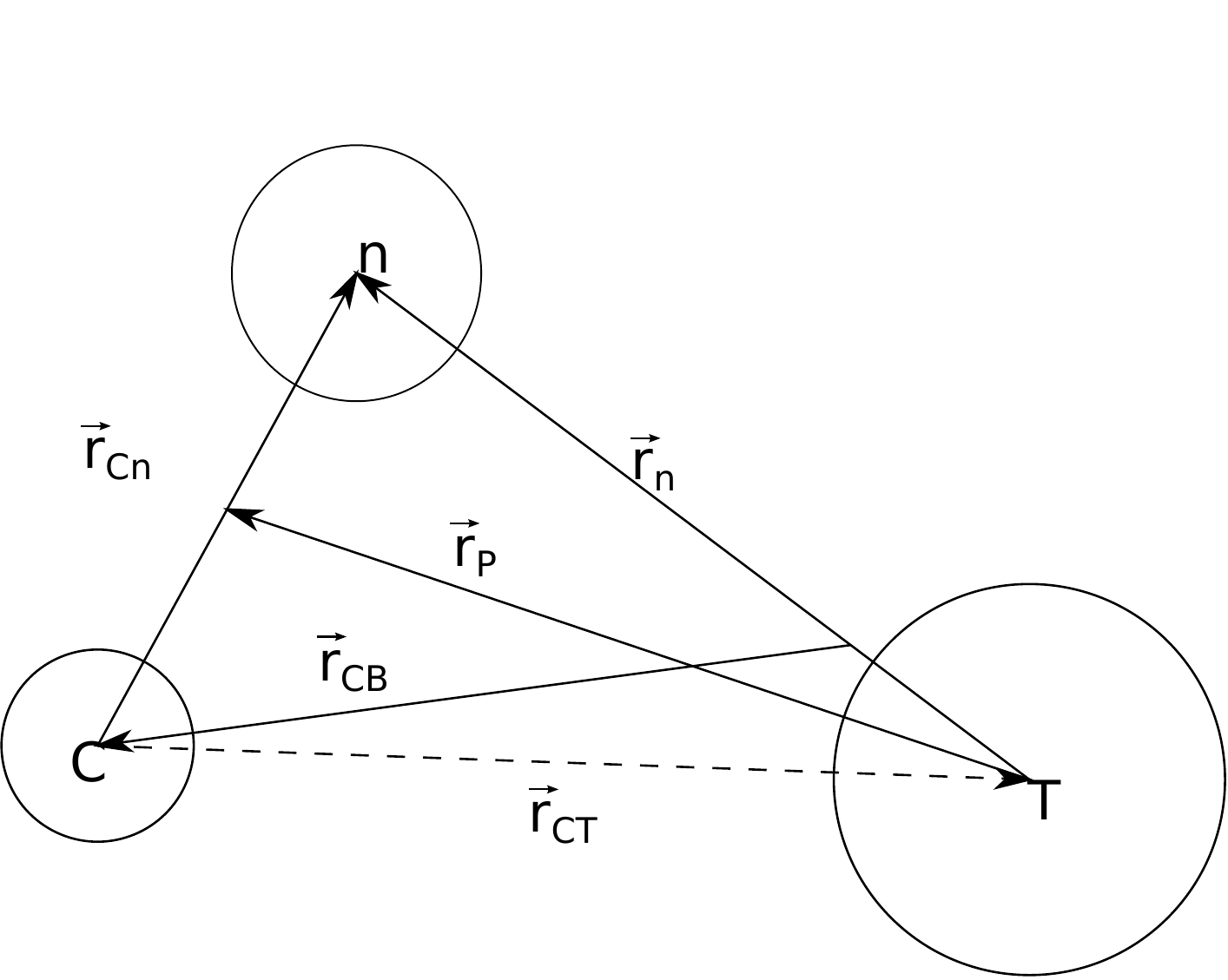}} \par}
\caption{\label{fig:IAV_coordinates}Coordinates used in the  models described in this work.}
\end{center}
\end{figure}

This process is   described by the effective Hamiltonian
\begin{equation}
\label{eq:IAV_H}
H=K+V_{C n}+U_{C T}\left(\mathbf{r}_{CT}\right)+H_{T}(\xi)+V_{n T}\left(\xi, \mathbf{r}_{n}\right),
\end{equation}
where $K$ is the total kinetic energy operator, $V_{Cn}$ is 
the interaction binding the two clusters $C$ and $n$ in the 
initial composite nucleus $P$, $H_T(\xi)$ is the Hamiltonian 
of the target nucleus (with $\xi$ denoting its internal 
coordinates), and $V_{nT}$ and $U_{CT}$ are the fragment-target interactions. The relevant coordinates are depicted in Fig.~\ref{fig:IAV_coordinates}.

In writing the Hamiltonian of the system in the form~(\ref{eq:IAV_H}) we make a clear distinction between the 
two cluster constituents; the interaction of the fragment 
$C$, the one that is assumed to be observed in the 
experiment, with the target, is described with a (complex) optical potential.
Inclusive breakup processes arising from this interaction
(e.g., target excitation) are included only effectively
through $U_{CT}$. The particle $C$ is said to act as 
spectator. On the other hand, the interaction of the particle $n$ with the target retains the dependence of the target degrees of freedom ($\xi$). In the following this dependence is released by the choice of a nucleon-target optical potential.

Starting from the  Hamiltonian of Eq.~(\ref{eq:IAV_H}) IAV derived the 
following expression for the double differential cross 
section for the inclusive breakup with respect to the angle 
and energy of the $C$ fragments:


\begin{eqnarray}
\label{eq:iav}
\frac{d^2\sigma}{dE_C d\Omega_C} =  \frac{2\pi}{\hbar v_P}\rho_C(E_C)\Big( \rho_n(E_n)\int |\langle \chi_n^{(-)} | \mathcal{S}_n  \rangle|^2 d\Omega_n  
- \langle G_n\mathcal{S}_n | W_n| G_n \mathcal{S}_n \rangle /\pi \Big)
\end{eqnarray}
where $v_P$ is the projectile-target relative velocity, $\rho_C(E_b) =
k_C\mu_C/((2\pi)^3\hbar^2)$ and $\rho_n(E_n) =
k_n\mu_n/((2\pi)^3\hbar^2)$ are the density of states for the particle $C$ and $n$ respectively, $W_n$ is the imaginary part of the optical potential $U_n$, obtained from the optical reduction of $V_{n T}\left(\xi, \mathbf{r}_{n}\right)$ which 
describes $n + T$ elastic scattering, $G_n= {1}/({E_n^{(+)}-U_n-T_n })$
is the Green's function for the nucleon-target channel, $E_n$ is the energy in the $n+T$ channel which satisfies the energy conservation, $\chi_n$ is the distorted-wave in this channel,
and $\mathcal{S}_n$ is the source term which takes the form 
\begin{equation}
\label{eq:source}
\mathcal{S}_n(\mathbf{r}_n) = \langle \mathbf{r}_n\chi_C^{(-)} |V_{post} |\Psi^{3b(+)} \rangle ,
\end{equation}
$\chi_C^{(-)}$is the distorted-wave describing the scattering of $C$ in the final channel with respect to the $n + T$ subsystem, and $V_{post} = V_{Cn} + U_{CT} - U_C$ (with $U_C$ the
optical potential in the final channel) is the post form transition
operator.
One should note that there is a natural separation between the elastic part and nonelastic part in Eq.~(\ref{eq:iav}), the first part corresponds to the elastic interaction between the $n$ and $T$ which is called elastic breakup (EBU), whereas the second term accounts for the cross section of the nonelastic process named nonelastic breakup (NEB). In this paper we shall discuss mainly the second NEB term. We show also the EBU term because of a short comparison we will make later with the corresponding term in the STC formalism.
There are various ways to approximate the
 three-body
wave function appearing in the source term of Eq.~(\ref{eq:source}). 
A simple choice which was implemented in \cite{jinme} and it allows comparison with both the HM and STC methods, is to use
the distorted-wave Born approximation (DWBA), i.e., $\Psi^{3b(+)}= \chi^{(+)}_{PT}(\mathbf{r}_P)\phi_P(\mathbf{r}_{Cn})$, where $\chi^{(+)}_{PT}(\mathbf{r}_P)$ is a distorted wave describing
$P + T$ elastic scattering and $\phi_P(\mathbf{r}_{Cn})$ is the projectile ground-state wave function.

Thus finally the NEB term can be written in a more compact form, which is useful for the comparisons we shall make in the next sections, as:

\begin{eqnarray}
\label{eq:iavc}
\frac{d^2\sigma^{NEB}}{dE_C d\Omega_C} &=& - \frac{2}{\hbar v_P}\rho_C(E_C) \langle \psi | W_n| \psi \rangle 
\end{eqnarray}
where $$\psi(\mathbf{r}_n)=G_n\mathcal{S}_n$$ and $G_n$ and $S_n$ are the Green function and source term discussed above.

\subsection{From IAV to HM and the eikonal approximation}

HM model \cite{Hussein:1985} differs from IAV \cite{IAV85} in the choice of the  wave function $\psi$ in Eq.(\ref{eq:iavc}) which they take as 
\begin{equation}\psi(\mathbf{r}_n)=\langle\chi{_{PT}}^{(-)}|\chi{_C}^{(+)}\phi_P(\mathbf{r}_{Cn})\rangle\label{HMwf}.\end{equation}

A big step forward in the understanding of the QM  models of breakup came about when HM made the hypothesis  in the formalism that 
the core and the nucleon breakup could be decoupled by summing the core-target optical potential to the nucleon-target optical potential:
 $$U^{PT}=U^{CT}+U^{nT}$$ and then assumed eikonal wave functions for the distorted waves in the entrance and exit channel. In this way the entrance wave function factorises into the core and neutron  wave functions  $\chi{_{PT}}^{(+)}=\chi{_{C}}^{(+)}\chi{_{n}}^{(+)}$  such that the two eikonal phases of the core in the entrance and exit channel sum-up providing the core-target S-matrix. Then  in the calculation of Eq.(\ref{HMwf}) the eikonal S-matrix of the core wave function factorises from the S-matrix of the nucleon wave function. This is indeed from the physics point of view similar to the starting point of the eikonal model of breakup of Yabana et al. and Hencken et al. \cite{YABANA,Hencken:1996} as summarised by Eq.(\ref{e1}) below and it is straightforward to obtain the final cross section formulae, Eqs.(\ref{ek34},\ref{ek35}). The great merit of the HM paper was then to make it clear what the core spectator model would be in practice and that the n-target S-matrix could safely be calculated on shell.
 
 Then  the equation for the eikonal "entrance channel wave function"  in the  HM paper, Eq.(4.9), after the   shift of variable $\br_{Cn}=\br_{CT}-\br_n$
 becomes
 
 \begin{equation}\psi_{eik}(\mathbf{r}_{Cn},{\bf k})
=\int d^{2}\mathbf{{r_P}_{\perp}}~e^{-i\mathbf{K}_{\perp}\cdot \mathbf{r_P}_{\perp}}~e^{i\mathbf{k}\cdot \mathbf{r}_{Cn}}
  {\bf S}_{c}\left(  \mathbf{b}_{c}\right) e^{  -\frac{i}{\hbar v}\int_{-\infty}^z d z ^{\prime} U^{nT} \left(
\mathbf{b,} z^{\prime} \right )} \phi_{0}\left(\mathbf{r}_{Cn}\right).  \label{HMe7}%
\end{equation}
which is interesting to compare  to the integrand of Eq.(\ref{e7}) in Appendix A, if the final nucleon wave function in Eq.(\ref{e7}) is chosen as a plane wave $\phi_{\mathbf{k}}^{\ast}=e^{-i\mathbf{k}\cdot \mathbf{r}_{Cn} }$. This shows that  with    the HM choice Eq.(\ref{HMwf})  of the  wave function $\psi$ in Eq.(\ref{eq:iavc}) and in the  the eikonal limit, the entrance channel was function  contains  almost the same factors  as  the integrand of the transition amplitude Eq.(\ref{e7})  before integration on the nucleon-core or nucleon-target impact parameter. Note that for final plane waves the order of integration of the  two integrals on the impact parameters of nucleon and core in Eq.(\ref{e7}) can be inverted. When inserted in Eq.(\ref{eq:iavc}) the wave function Eq.(\ref{HMe7})
 leads to the cross section for stripping because the eikonal phase for the nucleon is completed in the sense that one gets an integral between $-\infty$ and $+\infty$
 which leads to the nucleon-target  S-matrix.
The differential cross section Eq.(\ref{eq:iavc}) in the eikonal HM model  then reads:

\begin{equation}
\label{eq:EHM}
\frac{d^2\sigma}{dE_C d\Omega_C} =  \frac{2\pi}{\hbar v_P}\rho_C(E_C)\int ~d\mathbf{r}_{Cn} ~{\psi_{eik}}^{\ast}({\mathbf{r}_{Cn}},{\bf k}) W^{nT}\psi_{eik}({\mathbf{r}_{Cn}},{\bf k})\end{equation}


\subsection{STC model}

The semiclassical STC model \cite{bb,bb1,67,ppnp} is a generalization to final unbound states of the transfer between bound states model of Brink and collaborators \cite{hasan,Hasan_1979,Monaco_1985,27}. Semiclassical methods were very popular in the '70s as a substitute to full DWBA calculations which, as mentioned in the Introduction, were very lengthy  and computationally expensive. Transfer reactions were a common tool to study single particle  characteristics, in particular occupation probabilities but often  theoretical calculations  gave cross sections much larger than the data and spectroscopic factors different from the shell model  values~\cite{Winfield85,Winfield89,Pieper78,Olmer78}. Then an attempt to disentangle the content and the ingredients of the DWBA approach via semiclassical methods which could provide analytical expressions for the cross sections. The procedure followed was first to choose a  WKB wave functions for the distorted waves, similarly to HM \cite{Hussein:1985}, then the standard reduction of the three dimensional integral for the transfer form factor to a surface integral, similarly to what is done in Refs.~\cite{PAMPUS1978141,BAUR1984333,PhysRevC.84.044616,PhysRevC.89.054605,SD}. Finally analytical Hankel functions were used for the initial and final states on the surface between the two nuclei. The method is valid for peripheral reactions based on the core spectator model as mentioned above.
The use of the Hankel function, which is the  asymptotic form of the initial state wave function is an additional approximation in the STC which is not present in the HM formalism.
However the advantage is  that it allows the calculations to be carried on analytically up to the end allowing for a transparent interpretation of the formalism and of its results, therefore we remind in the following a few steps leading to the final probability and cross section formulae Eqs.~(\ref{dpde}) and~(\ref{totx}). The full  formal derivation is given in the Appendices B and C.

 The semiclassical transfer to the continuum amplitude \cite{bb}  is: 
\begin{equation}
A_{fi}=\frac{1}{ i\hbar}
\int_{-\infty}^{\infty}dt\langle\phi_{f} ({\bf r_n})|U_{nT}({\bf r_n})|\phi_{i}({\bf r_n-R}(t))\rangle e^{-i\bar\omega}\label{1}
\end{equation}
with $\bar\omega=(\omega t-mvz/\hbar)$. 
The time dependent nucleon initial and final wave functions in their respective reference frames are  \begin{equation}\psi_{i,f}({\bf r_n},t)=\phi_{i,f} ({\bf r_n})e^{-i\varepsilon_{i,f} t/ \hbar}\label{wf1}\end{equation}
Initial and final radial wave functions are taken as Hankel functions according to \cite{bb} such that we have for the
initial state:
\begin{eqnarray}
\phi_{l_i}({\bf r_{Cn}})=-C_ii^l\gamma h^{(1)}_{l_i}(i\gamma_i r_{Cn})Y_{l_i,m_i}(\Omega_i).  
\label{wf}\end{eqnarray}
And for the 	final continuum state a  scattering wave function defined as: 
				
\begin{eqnarray}
\phi_{l_f}({\bf r_n})=C_f k_f\frac{i}{2}(h^{(+)}_{l_f}(k_fr_n)- S_{l_f} h^{(-)}_{l_f}(k_fr_n))Y_{l_f,m_f}(\Omega_f),
\end{eqnarray}
$\ S_{l_f} (\varepsilon_f)$ is the  n-target  S-matrix at energy $\varepsilon_f$.\vskip .5cm

For proton breakup we use the same type of wave functions. For the initial state we first calculate the exact proton  bound state wave function and then we fit to it a neutron wave function. Finally  we use such a neutron wave function and the corresponding, effective separation energy.  This method was checked and found very accurate in Ref.\cite{Ravinder}.

Using the above definitions in the amplitude Eq.(\ref{1}) and then taking the modulus square, the breakup probability reads:
\begin{eqnarray} 
\frac{dP_{-n} }{d\varepsilon_f}\approx \frac{1}{2}\Sigma_{j_f}(2j_f+1)|1- S_{j_f} |^2
(1+R_{if})
\left [\frac{\hbar}{mv}\right ]^2
\frac{m}{\hbar^2 k_f}
|C_i|^2 
\frac{e^{-2\eta b_c}}{2\eta b_c}
M_{l_fl_i}, \label{dpde}
\end{eqnarray} 

Note that up to this stage the  practical choice of the n-target S-matrix has not been made yet, we have only required that the final n-target state be a scattering state with respect to the target describing  the n-target  FSI. The above formula is very general and as such it contains all possibilities discussed in the models of the Introduction, when the interest is not on the n-core FSI. As it was shown in Ref.\cite{bb},  the following choices can be made     for  the model potential and its possible energy dependence, describing  different physical situations:
\begin{itemize}
\item An energy {\it independent} real potential and just one angular momentum state $j_f$  would describe  a low energy single particle resonance in the target holding the correct normalization (c.f. Eq.(3.2) of  \cite{bb}).
\item An energy {\it independent} real potential and the sum over final spin states of the target  $j_f$ would describe elastic breakup including the sum of narrow resonances.
\item An energy {\it dependent} optical potential would describe compound nucleus resonances, both at low and high n-target energies. At low energies the resonances will be narrow, at high energies they will be broad and overlapping. In this case one needs to keep the sum over final states in Eq.(\ref{dpde}) and the formula needs energy averaging, according to the optical model, under the hypothesis that only the term $|1-S|^2$ has a strong energy dependence.
 Thus one obtains two terms representing the so-called elastic and inelastic breakup. \begin{equation}
\frac{dP_{-n} }{d\varepsilon_f}\approx \frac{1}{2}\Sigma_{j_f}(2j_f+1)(|1- \bar S_{j_f} |^2+1- |\bar S_{j_f} |^2)B_{if}\label{dpde1}\end{equation}
where $B_{if}$ contains all smoothly varying energy dependent factors of Eq.(\ref{dpde}).

Then apart from the choice of the potential one can further discuss the energy averaging procedure and the method to calculate the S-matrix. 
\item Suppose the incident energy is high enough for the matching conditions to favour high energy states of the nucleon in the continuum in which the potential can be considered smoothly varying and no resonances are present. In this situation an eikonal model calculation of the S-matrix can be acceptable and thus taking the classical limit in Eq.(\ref{dpde}) $\Sigma_{j_f}\to \int d^2\bf{b_n}$ the eikonal elastic breakup formula could be obtained but only in presence of a real potential.

\item If absorption is present then  the energy averaging is to be done according to Eq.(\ref{dpde1}) but still the S-matrix can be calculated in the eikonal approximation. In this way elastic and inelastic scattering would be obtained consistently in the eikonal approximation. From the nucleon energy distribution one can get the nucleon momentum distribution by using Eq.(\ref{k12}) below and the appropriate Jacobian. Note that in this case the eikonal limit taken from Eq.(\ref{dpde}) would satisfy the kinematical condition $\varepsilon_f^{min} =0$ translating into $k_1^{min}=-(\varepsilon_i+\frac{1}{2}mv^2)/(\hbar v)$.
\end{itemize}

To summarise: in Eq.(\ref{dpde1}) $\bar S_{j_f}$ are nucleon-target S-matrices calculated for each nucleon final energy according to the optical model in an energy dependent optical potential, including the spin-orbit term of the nucleon-target optical potential. The sum of the two terms $(|1-\bar S_{j_f} |^2+1-|\bar S_{j_f} |^2)$ is obtained automatically \cite{bb} as a result of using an unitary  energy averaged optical model S-matrix thus  including  non elastic and elastic breakup (stripping and diffraction). These two terms correspond to the first and second term   of  Eq.(\ref{eq:iav}). The sum over partial waves in Eq.(\ref{dpde}) is  a sum over total   nucleon-target angular momenta. $C_i$ is the initial wave-function asymptotic normalization constant. It is obtained as the ratio between the  numerically calculated single particle wave function and the Hankel function. The form factor  $\frac{e^{-2\eta b_c}}{2\eta b_c}$is due to the combined effects of the initial and final wave-function Fourier transforms,  while $M_{l_fl_i}$ is due to the overlap of the angular parts. $R_{if}$ are spin-coupling coefficients. Further definitions and discussion can be found in Refs.\cite{initst,ppnp} and in  Appendix B.

In the core spectator model the  breakup  cross section is obtained  by integrating the differential breakup probability  on the core-target impact parameter by  weighting it with the probability $|S_{ct}(b_c)|^2$ that the  measured core
has survived  "intact" the scattering. This term in the HM model comes from the eikonal choice of the projectile and core distorted waves exactly as in the STC \cite{hasan,Hasan_1979}. Finally if a shell model Woods-Saxon wave functions is used one multiplies  by $C^2S$ the spectroscopic factor of the initial state

\begin{equation}\frac{d\sigma _{STC}}{d\zeta}=C^2S
   \int_0^{\infty} d{\bf b_c}   |S_{ct}(b_c)|^2 {\frac {d P_{-n}(b_c)} {d\zeta} }.\label{totx}\end{equation}
In Eq.(\ref{totx}) the variable $\zeta$  can be the nucleon final energy in the continuum $\varepsilon_f$ as in Eq.(\ref{dpde1}) and/or the nucleon momentum with respect to the core or target given in Eq.(\ref{k12}) below, or
by using 4-energy momentum conservation (see for example \cite{firk}) and the relative Jacobian,  the differential $d{\sigma_{STC}/ {d\zeta}}$ cross section becomes directly comparable to the measured momentum distributions function of P$_{//}$ the core parallel momentum. 

 This formalism does not include Coulomb recoil effects of the core because it does not distinguish the center of mass of the core-target system from the center of mass of the projectile-target, c.f.  the coordinates  $r_{CT}$ and  $r_{P}$ in Fig.\ref{fig:IAV_coordinates}. Core recoil effects give rise to the so called Coulomb breakup which is important for heavy targets and very weakly bound initial states. It can be treated together with the nuclear breakup according to \cite{Ravinder,jer1,jer2,Alvaro1}. On the other hand, STC contains the energy recoil effect of the nucleon via the definitions of 
 
 \begin{equation}k_{1,2}=(\varepsilon_f-\varepsilon_i\mp\frac{1}{2}mv^2)/(\hbar
v)\label{k12}\end{equation} 
which can be interpreted as the z-components of the nucleon momentum in the initial (core) and final (target)
reference frames respectively, see Appendix B. Those can be sampled in a breakup  reaction according to the kinematical constraints. 

\section {Comparison of the models}
\begin{figure}[tb]
\begin{center}
 {\centering \resizebox*{0.75\columnwidth}{!}{\includegraphics{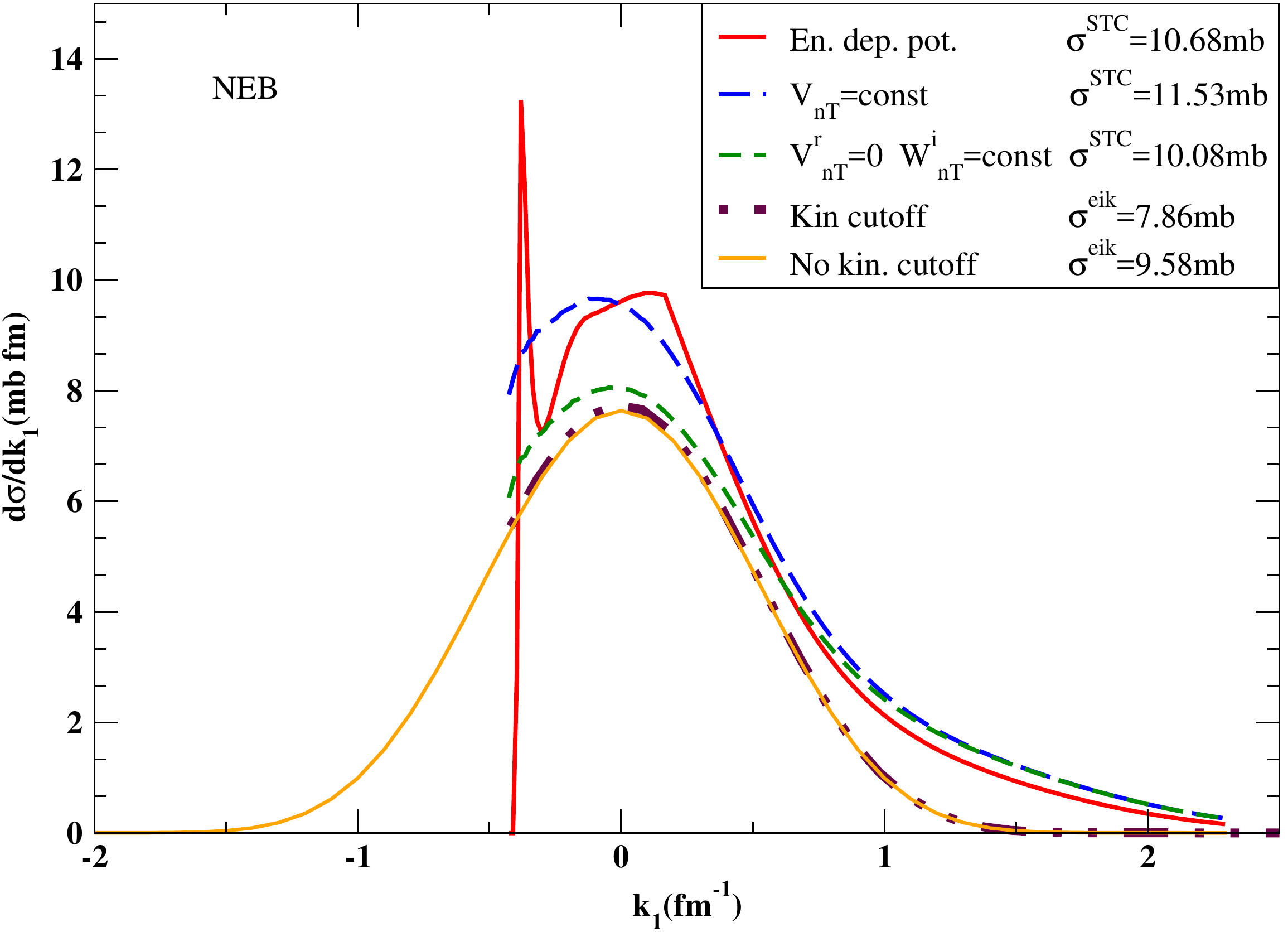}} \par}
\caption{\label{fig:k1dis} NEB (stripping) neutron momentum distributions in the n-Core reference frame for the reaction $^9$Be($^{14}$O,$^{13}$O,)X at 53 A.MeV incident energy.  The full red curve represents the STC model result, the blue long dashed curve is  obtained by setting the strength of the n-T optical potential constant and equal to the value at 53 MeV,  the green short dashed curve is obtained by setting the real part of the n-T potential equal to zero. The orange thin full curve shows the standard eikonal calculation, while the brown double-dotted-dashed curve shows the eikonal results in which the kinematics cutoff has been implemented. See text for more details.}
\end{center}
\end{figure}

\begin{figure}[tb]
\begin{center}
 {\centering \resizebox*{0.75\columnwidth}{!}{\includegraphics{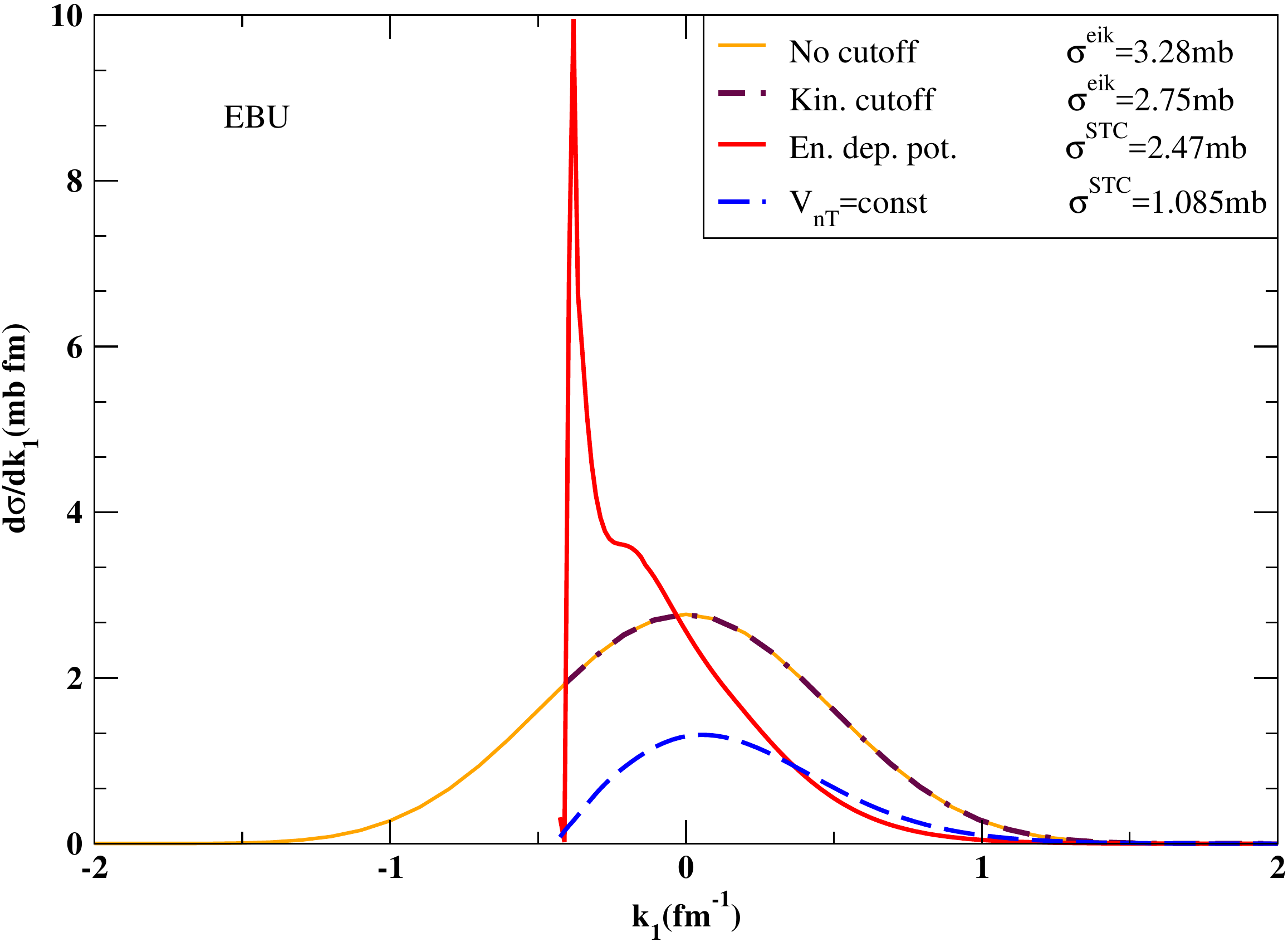}} \par}
\caption{\label{fig:k1diff} Same reaction and same notation as  in Fig. \ref{fig:k1dis} but for EBU (diffraction) neutron momentum distributions in the n-Core reference frame. See text for more details.}
\end{center}
\end{figure}

\begin{figure}[tb]
\begin{center}
 {\centering \resizebox*{0.75\columnwidth}{!}{\includegraphics{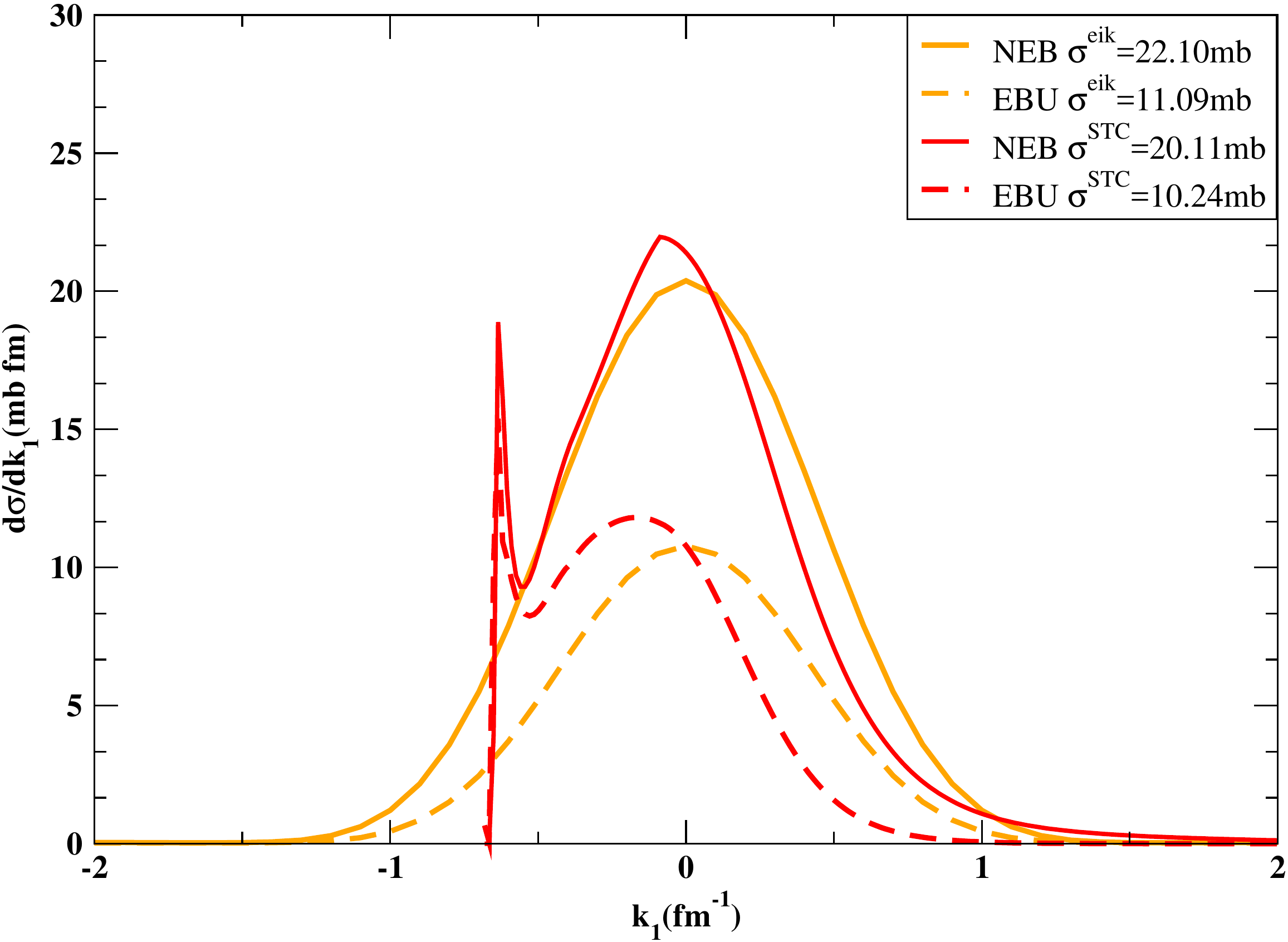}} \par}
\caption{\label{fig:k1disp}  Proton breakup  momentum distributions in the n-Core reference frame for the reaction $^9$Be($^{14}$O,$^{13}$N,)X at 53 A.MeV. Red  and orange curves are from the STC and eikonal models respectively. Full lines for NEB, dashed lines for EBU.}
\end{center}
\end{figure}

In this section we make a critical comparison of the IAV, HM and STC model described before.
In general it is clear that any proper theory of, at least, one nucleon breakup, should  consistently provide the elastic and inelastic breakup terms. It should also be general enough to be applicable  in a large range of incident energies, to any projectile-target combination and with no constraint on the initial nucleon separation energy and/or the number of possible projectile excited states. The IAV model satisfies those requirements while the eikonal HM model is a high energy model.

The STC formalism described in the previous section partially satisfies all the above requirements.   It treats on the same footing both the elastic and inelastic breakup. It can be applied with energy dependent or independent potentials, real and/or complex  and with  different energy averaging procedures. Besides,  fully analytical formulae can be obtained which help disentangling the complicated physics involved.  Both the eikonal and STC formalism need an energy averaging 
with respect to the original IAV model but while the eikonal averages over the whole energy range spanned by the nucleon in the continuum, the STC averages the nucleon-target interaction over small intervals, according to the optical potential that fits the nucleon-target free particle cross section. The IAV model when implemented with an energy  dependent optical potential contains the same energy averaging. 
The STC can treat also cases in which for a heavy target, the n-target system has high lying resonances in the continuum  
\cite{Highensp,Tina}. At present, due to the high number of partial waves required,  the numerical implementation of the NEB by the IAV model can treat only low-energy
and small projectile transfer to target resonances. The eikonal model cannot treat resonances. Finally it should be noted that the STC does not need to average the initial and final n-core and n-target momenta, thus naturally satisfying energy conservation for the nucleon, even when the S-matrix is calculated in the eikonal model. 
 However in the STC the core-target S-matrix  is treated in the eikonal approximation similarly to  HM. For this reason the STC model can be defined as an intermediate model, because it treats the core-target relative motion in the eikonal approximation while the n-target motion and interaction is treated quantum mechanically. Energy conservation and correct kinematics within the eikonal model have been discussed also by Fujita and H\"ufner \cite{FH} and Ogata et al. \cite{ogata2015}.

Recently it has been shown that the IAV and STC methods lead to very close results \cite{jinme}. However from the formal point of view they look quite different if one compares Eq.(\ref{eq:iav}) with Eqs.(\ref{dpde})
and (\ref{totx}).  The main differences are that in the IAV method the core-target $S$-matrix is included in the source term Eq.(\ref{eq:source}) and and n-target $S$-matrix is considered as solution of an  inhomogeneous equation while in STC they are solutions of homogeneous equations. In practice this means that the STC method decouples the core-target scattering from the nucleon-target scattering, considering them as independent. This  corresponds to consider off-shell effects negligible and thus it calculates  on-shell S-matrices. The historical origin of this difference lies in the fact that IAV approach originated as a method to calculate light nuclei ($d$) breakup at low energy while STC was developed to treat heavy-ion reactions at intermediate to high energies where surface approximation and thus decoupling of core-valence-particle degrees of freedom
appear as the natural choice. Indeed one can see that such a decoupling could be done also in the IAV model by approximating the distorted wave for the projectile-target relative motion  in the same way as done by HM, with the core-target distorted wave in the entrance channel multiplied by the nucleon distorted wave  $\chi{_{PT}}^{(+)}\approx \chi{_C}^{(+)}\chi{_n}^{(+)}$. In fact, in the case of heavy-ions  the projectile and core have nearly the same mass,  the core is much heavier than the nucleon and the relative motion trajectory of the projectile and core are nearly the same, as assumed by the STC model. This again can be called the no-recoil approximation.

\subsection {Numerical results}

In this section we present results of numerical calculations for the neutron and proton breakup reactions

 \noindent {$^9$Be($^{14}$O,$^{13}$O,)X, $^9$Be($^{14}$O,$^{13}$N,)X at 53 A.MeV.  The neutron breakup reaction has become a test case for breakup models since the data and theoretical analysis  by STC were presented in \cite{Flavigny:2012}. In fact the experimental final core spectrum  has shown the strongest deviation from the eikonal predictions, among the cases in literature, highlitening the role of kinematics and FSI effects. The reason is that the initial neutron separation energy is 23.2 MeV while the incident energy per nucleon was  53.2A.MeV. Thus the medium available energy in the neutron continuum was only about 30MeV. Recently the model and calculations  by STC have been benchmarked by comparing them with the IAV model \cite{jinme}, resulting in excellent agreement. The goal  of this section is to compare the results of the STC with the eikonal model and to clarify the correct procedure to perform such comparison.}
 \indent 
 \begin{itemize}
 \item First it must be stressed, as argued earlier on in this paper,  that at  energies above the threshold for the first excited state in a n-target (n-T) interaction,  the optical model  requires an energy dependent strength  for the complex potential. The IAV model and STC require the same for the potential providing the final state of the breakup nucleon. The eikonal model of breakup requires a complex potential but averages over the whole energy range, which means its implementation is with the n-T potential calculated at the incident energy per nucleon.
 \item The IAV and STC contain the correct kinematics and energy conservation. The standard eikonal model of breakup neglects both, however it was shown in \cite{geopap} that the kinematical cutoff could be easily implemented in the eikonal model. Effects of this correction, linking the nucleon separation energy with the incident energy were numerically predicted in \cite{geopap} and verified in \cite{Flavigny:2012,jinme}.
 \end{itemize}
 
 In Figs.\ref{fig:k1dis},\ref{fig:k1diff} the results for the neutron NEB and EBU (stripping and diffraction) are presented respectively. Momentum distributions of the cross section with respect to $k_1$, the z-component of  the neutron momentum in the core reference frame are  shown. In this representation the eikonal model supposes that the measured distribution represents the neutron initial bound state momentum distribution. It is understood that the z-axis is the relative motion velocity axis in the reaction. We have used the same initial state parameters as in \cite{jinme}, the same energy dependent n-T optical potential from \cite{bobme} and the single folding procedure from \cite{BCC1,BCC,imane} to calculate the imaginary core-target potential which gives the core-target S-matrix. The following analysis has partially been prompted by discussions with the authors of Ref.\cite{jinall}  and the interpretation of results therein. In particular we are trying to understand the limits of applicability of the eikonal model and which is the accuracy we want to set on a given formalism to be considered reliable. In Ref.\cite{jinall} 50A.MeV is considered as an energy already high enough to justify the use of the eikonal approximation.  Contrary to that, we will show in the following that there are non negligible differences between a QM or semiclassical calculation and the eikonal result.
 
 In both figures the notation is the same and in the legends the integrated cross sections corresponding to each curve are given. 
 The full red curves are obtained by the STC model, the blue long dashed curves are obtained setting the strength of the real and imaginary parts of the  n-T optical potential constant and equal to the value at 53 MeV, as it is done in the eikonal calculation, the green short dashed curves are obtained by setting the real part of the n-T potential equal to zero, as  in the stripping calculation by the eikonal model. On the other hand the orange thin full curves show the standard eikonal calculation, while the brown double-dotted-dashed curves show the eikonal results in which the kinematical cutoff has been implemented.
 
 The correct comparison  requires that each method be implemented according to the physics that it represents, thus we should look first of all to the red curves vs. the orange curves corresponding to the STC and eikonal. It is clear that the curves are rather different and the integrated cross sections differ by about 20\% for NEB. 
 
 One might naively imagine that the differences are due to the different potentials and/or to the kinematical cut off. To clarify this point we look at the blue long dashed curves and at the green short curve. Both curves and integrated cross sections are different from the standard  STC and  eikonal results. Thus there is no way to reconcile the STC, which is equivalent to a QM method such as the IAV, with the eikonal. One might also wonder on the role of the kinematical cutoff. Even the kinematically-corrected eikonal (double-dotted-dashed brown curve) does not fit any of the other spectra, nor it does the integrated cross section.
 Thus it is clear that for NEB cross sections a typical 20\% difference is expected  between  STC results and the eikonal. The spectra look also different but probably in general not distinguishable in comparison to the data, apart from the kinematical cutoff.
 
The differences in shape are even   more noticeable for the spectra of the EBU, Fig.\ref{fig:k1diff}  but  the STC results and the eikonal results can be reconciled for the total cross section if one implements the kinematical cutoff in the eikonal.  Otherwise the cross sections differ by about 30\% the eikonal giving the larger value. There is no green short dashed curve in this case because  for  EBU the real potential cannot be set to zero.   In this  case the eikonal cross section is larger than the STC. 
 Finally summing up NEB and EBU as given by their correct implementations, we obtain $\sigma_{tot}$=13.15mb from STC (red curves) and
 $\sigma_{tot}$=12.86mb (orange curves) from eikonal. 
 
 We note that the STC results show a spike close to the neutron threshold. It is due to the excitation of low energy resonances in the n+$^9$Be systems discussed in Ref.\cite{Flavigny:2012}. They are seen in the  n+$^9$Be data and reproduced by our n-T optical potential \cite{bobme}. Their presence in the STC results is a proof that this method can indeed be used to study transfer  to target resonances as mentioned in Sec. 2.3 while the eikonal can be used only to get information on the projectile initial state.

For completeness in Fig.\ref{fig:k1disp} the spectra of the proton breakup reaction are presented.  The red curves represent the SCT, full line is the NEB  while dashed line is the EBU. The orange curves give the eikonal results. In this case kinematical effects are less important because the proton separation energy is only 4.63MeV.
In fact the maxima of the spectra are closer to each other and to the $k_1=0$ point where the maximum would be expected if the  momentum distributions after breakup represented exactly the nucleon  momentum distributions in the initial states. In this case the eikonal gives larger cross sections than the STC, the difference being 20\% for NEB and 10\% for EBU. The summed cross sections are $\sigma_{tot}$=30.35mb from STC (red curves) and
 $\sigma_{tot}$=33.19mb (orange curves) from eikonal. 
 
 The conclusion of this section is that in both neutron and proton breakup case and from a very deep and less deep initial bound state the difference between the eikonal results and the STC results for the total  cross sections are about 20\% in agreement with our previous findings \cite{geopap}. However the shape of the spectra can be very different  demonstrating the effects of kinematics and FSI with the target. It appears that the use of an energy dependent FSI is more consistent with a QM model. 
 
\section{Conclusions}

In this paper we have  reviewed  some of the models of breakup that at present are able to calculate heavy ion reactions and have been applied to exotic nuclei studies. We have concentrated in particular on the inelastic breakup mechanism. In principle the IAV is the most accurate method on the market and it allows also the use of FSI potentials with energy dependent strengths \cite{jinme}. It is also very successful in describing transfer to continuum at low energy \cite{Jin15,Jin18b,Jin17,Jin15b,Jin18}. At the moment it cannot deal with high number of partial waves, which means it cannot calculate  breakup on heavy targets. Also in the version with DWBA final wave functions it cannot calculate halo breakup because the process is not perturbative.  However this is just a technical problem which hopefully will be solved in a near future. IAV integrated cross sections  and the shape of the spectra from breakup on a small $^9$Be target \cite{jinme} at intermediate energies are in agreement with STC values thus providing a cross-check for both models.

The eikonal model for stripping originated from HM seminal paper  \cite{Hussein:1985}. A derivation of  the eikonal diffraction term   from IAV  does not exist   but  from the STC elastic breakup formula taking the eikonal limit one can obtain an expression consistent with the diffraction no recoil approximation \cite{Angela2018}. The eikonal model obviously cannot describe transfer to resonances or detailed structures of inclusive spectra originating from FSI effects and/or kinematics. This is  because it assumes an energy averaging procedure over the whole energy spectrum. However kinematics can be implemented in it at various levels \cite{FH,geopap,ogata2015}. Total cross sections differ typically by 20\% with respect to QM models but this effect is not systematic.

The STC originated \cite{bb,bb1} from the need to interpret the physical content of all QM models available in the '70s and '80s which we have reviewed in the Introduction. Its derivation is completely different from them but it assumes the core spectator model as in the HM model and the method is basically an extension of previous semiclassical transfer models to the case of a continuum final state. In this sense it is quite similar to \cite{PAMPUS1978141,BAUR1984333}. Having been implemented with energy dependent FSI potentials it has been able to reproduce a quite large number of different reaction data over more than thirty years. Besides it is numerically rather trivial which allows for a multitude of applications. It is accurate to describe low and high lying resonances \cite{Highensp,Tina} as well as the three-body elastic breakup background \cite{AB1} and breakup from weakly and strongly bound initial states. An interesting result of the earlier applications to heavy targets such as $^{208}$Pb was to show that single particle states in the continuum were reproduced in position and width thanks to the use of the accurare Mahaux and Sartor n-Pb potential \cite{Mahaux:91}. Thus it is clear that the early problematic distinction between breakup fusion  in the elastic and/or  inelastic channel, the question of fluctuations in the cross section vs direct part, etc...they are all solved by the use of an accurate energy dependent FSI potential. This was true for lead and other heavy targets and it is true nowadays for the $^9$Be target. Furthermore  
there is no limitation on the initial state angular momentum that STC can treat because the formulae are analytical. In the Appendices B and C we provide the most formal derivation of it which allows, see  in particular Appendix C, a systematic inclusion of FSI to all orders and helps clarifying the content of other models.

To study properties of the last nucleon in  short-lived, exotic nuclei, one-nucleon breakup at intermediate energies has been used in the last fifteen years and has largely contributed to establishing the picture of shell structure away from stability by  extracting the spectroscopic factors~\cite{Macfarlane} for the initial state wave function from the comparison of experimental data to the reaction theory predictions. 
A recent compilation of experimental  integrated knockout cross sections at intermediate energies has shown a systematic trend when compared to theoretical calculations based on shell-model predictions for shell occupancy and eikonal approximation for the nucleon removal reactions~\cite{Tostevin_Gade}.  
However, this marked dependence does not seem to be supported by the results obtained with transfer reactions~\cite{Lee11,Flavigny13,Flavigny18} and quasifree
scattering with ($p$, $2p$), ($p$, $pn$), and ($e$, $e'p$) reactions \cite{PPNPall}. It is worth noting that the theoretical cross section depends on the description of the reaction mechanism but also on the choice of the initial state wave function. In this paper we have shown that detailed features of the spectra which are  not represented by the eikonal procedure are also somewhat washed out by the energy integration of QM results when the same FSI and initial wave functions are used. On the other hand strong dependence on the ingredients used in the calculations has been noticed and studied in Ref.\cite{AG}. Thinking of the $^{14}$O breakup discussed in this work  one might notice that a  different eikonal result  was quoted in Ref.\cite{Flavigny:2012} than the one given here, while the STC results are consistently the same. This is because in Ref.\cite{Flavigny:2012} the eikonal S-matrices were obtained with different FSI, although the initial state  wave function was the same as in the present work. It was noted in Ref.\cite{AG} that the strongest dependence in the eikonal results was from the wave function radius parameter choice rather than from the S-matrices calculations.

These discussions  are very important in view of the present-state-of-the-art of both experimental and theoretical knockout studies \cite{PPNPall} and our findings can open up new avenues to the theoretical understanding of nuclei with very unbalanced N/Z ratios.
In several  papers quoted in Ref. \cite{Tostevin_Gade}  the spectra are not shown and only the total cross sections are quoted. Suppose the results of our present paper were general. One would say that the eikonal  gives smaller values than a QM model for deeply  bound states and larger values for weakly bound states. Also the trend is not always the same for diffraction and stripping. Now while the cases studied here correspond to final cores ($^{13}$O and $^{13}$N) not having bound excited states, most of the results quoted in  \cite{Tostevin_Gade}  have to do with nuclei in which several core excited states contribute to the inclusive cross section. One can imagine that there could be a kind of compensation and the slope of the plot for the reduction factor, defined as the ration of $\sigma_{exp}/\sigma_{th}$ could be a combination of some systematic effect. Therefore a critical review of the plot would require a comparison of results from two models (eikonal and STC or IAV), nucleus by nucleus, state by state, differentiating EBU and NEB and always cross-checking the spectra.  If the results of  \cite{Tostevin_Gade}  would be confirmed, together with the recent work on knockout at the proton drip line \cite{boblast} the logical  conclusion would be that the shell model concepts have to be completely revised  at and across the driplines.
\section{Acknowledgements}

One of us  (AB) is grateful to M. G\'omez Ramos, J. G\'omez-Camacho, Jin Lei  and A. M. Moro for several discussions.


\begin{appendix}
\section{Reminder of eikonal formulae}

Following Refs. \cite{YABANA,Hencken:1996,BG} we consider a single-particle model for a 
halo nucleus and introduce the eikonal approximation to study its scattering on another target nucleus. The ground state is described by a wave function $\phi_{0}(\mathbf{r_{Cn}})$ which depends on the relative coordinate $\mathbf{r_{Cn}}$, Fig.\ref{fig:IAV_coordinates}. between
the nucleon and the core. After interacting with the target the eikonal
wave-function of the halo nucleus in its rest frame has the form

\begin{equation}
\Psi\left(  \mathbf{r}_{Cn}, \mathbf{r_{P}}\right)  =S_{n}\left(  \mathbf{b}_{n}\right)  S_{CT}\left(
\mathbf{b}_{c}\right)  \phi_{0}\left(  \mathbf{r}_{Cn}\right)  \label{e1}%
\end{equation}
where $\mathbf{r_P}$  and $\mathbf{r_{Cn}}$ are the coordinates of the center-of-mass of the projectile
consisting of the core plus one nucleon, and of the nucleon with respect to the core respectively, see Fig.\ref{fig:IAV_coordinates}. The vectors
\begin{equation}
\mathbf{b}_{n}=\mathbf{r_P}_{\perp}+\beta_{2}\mathbf{r_{Cn}}_{\perp}\qquad
\mathrm{and}\qquad\mathbf{b}_{c}=\mathbf{r_P}_{\perp}-\beta_{1}\mathbf{r_{Cn}}%
_{\perp} \label{e2}%
\end{equation}
are the impact parameters of the nucleon and the core with respect to the
target nucleus. Thus $\beta_{1}=m_{n}/m_{p}$, $\beta_{2}=m_{c}/m_{p}%
=1-\beta_{1}$, where $m_{n}$ is the nucleon mass, $m_{c}$ is the mass of the
projectile core and m$_{p}=m_{n}+m_{c}$ is the projectile mass. The two
profile functions $S_{n}$ and $S_{cT}$ are defined in terms of the
corresponding potentials by
\begin{equation}
S\left(  \mathbf{b}\right)  =\exp\left(  -\frac{i}{\hbar v}\int dzV\left(
\mathbf{b,}z\right)  \right)  \label{e3}%
\end{equation}
where $v$ is the beam velocity. The breakup amplitude generated from the
eikonal wave function (\ref{e1}) has a direct contribution from the
nucleon-target optical potential $V_{nT}$ represented by nucleon-target
profile function $S_{n}$ and a core-recoil contribution from the core-target
interaction $V_{CT}$ represented by the profile function $S_{CT}$ \cite{jer1,jer2}. The recoil
contribution depends on the ratio $\beta_{1}$ of the nucleon mass to the
projectile mass and goes to zero in the limit $\beta_{1}\rightarrow0$. The
potential $V_{CT}$ includes the core-target Coulomb potential and the real and
imaginary parts of the nuclear potential. The Coulomb part of $V_{CT}$ is
responsible for Coulomb breakup. Using the approximate form of  
 the wave function (\ref{e1}) with (\ref{e3}) implies the ''frozen halo'' approximation; the nucleon
velocity relative to the core in the projectile and in the final state is slow
compared with the incident velocity $v$.

\bigskip

The eikonal breakup amplitude is defined by \cite{Hencken:1996}
\begin{eqnarray}
A\left(  \mathbf{K,k}\right)  =\int d^{2}\mathbf{r_P}_{\perp}~e^{-i\mathbf{K_{\perp}\cdot r_P}%
_{\perp}} \int d^{3}\mathbf{r_{Cn}}~\phi_{\mathbf{k}}^{\ast}\left(  \mathbf{r}_{Cn}%
\right )  \left(  S_{C}\left(  \mathbf{b}_{C}\right)  S_{n}(\mathbf{b}%
_{n})-1\right)  \phi_{0}\left(\mathbf{r_{Cn}}\right).  \label{e4}%
\end{eqnarray}
The impact parameters $\mathbf{b}_{n}$ and $\mathbf{b}_{C}=\mathbf{b}%
_{n}+\mathbf{r}_{\perp}$ are defined in Eq.(\ref{e2}). The quantities $\left(
\mathbf{K,k}\right)  $ are the momenta conjugate to the coordinates $\left(
\mathbf{r_P,r_{Cn}}\right).$ They are related to the final momenta of the core,
nucleon and target by
\begin{equation}
\mathbf{k}_{C}=-\mathbf{k}+\beta_{2}\mathbf{K},\qquad\mathbf{k}_{n}%
=\mathbf{k}+\beta_{1}\mathbf{K,}\qquad\mathbf{k}_{T}=-\mathbf{K.} \label{e5}%
\end{equation}
The wave function $\phi_{\mathbf{k}}\left(  \mathbf{r}_{Cn}\right) $ is the final
continuum wave function of the nucleon relative to the core. The complete
differential cross-section is 
\begin{equation}
\frac{d\sigma}{d^{2}\mathbf{K}d^{3}\mathbf{k}}=\frac{1}{\left(  2\pi\right)
^{5}}\left|  A\left(  \mathbf{K,k}\right)  \right| ^{2} \label{e6}%
\end{equation}
Eq.(\ref{e4}) can also be written as%

\begin{eqnarray}
A\left(  \mathbf{K,k}\right)  =\int d^{2}\mathbf{r_P}_{\perp}~e^{-i\mathbf{K_{\perp}\cdot r_P}%
_{\perp}} \int d^{3}\mathbf{r_{Cn}}~\phi_{\mathbf{k}}^{\ast}\left(  \mathbf{r}_{Cn}%
\right)  S_{C}\left(  \mathbf{b}_{C}\right)  S_{n}(\mathbf{b}%
_{n})  \phi_{0}\left(\mathbf{r}_{Cn}\right).  \label{e7}%
\end{eqnarray}
because of the orthogonality of $\phi_{\mathbf{k}}\left(  \mathbf{r}_{Cn}\right) $
and $\phi_{0}\left(  \mathbf{r}_{Cn}\right)  $ (cf. Eq.(8) of Ref.\cite{Hencken:1996}).  Equations
(\ref{e7}) is a general eikonal expression which has been used in
\cite{Hencken:1996} and by many other authors.

Finally following the derivation in \cite{Angela2018} the EBU eikonal cross section in the no-recoil approximation 
 is
 \begin{equation}
\frac{d\sigma_{EBU}}{dk_1}=\int d^{2}\mathbf{b}_{C}~|S_{CT}\left(\mathbf 
{b}_{C}\right) |^2\int d^{2}\mathbf{r}_{{Cn}_{\perp}}~\left| 1-S_{n}\left( \mathbf{b}
_{n} \right )\right |^2  |{\tilde \phi}_{0}\left(  \mathbf{r}_{{Cn}_{\perp}},k_1\right)|^2,
\label{ek34}%
\end{equation}

and the NEB formula \cite{YABANA,Hencken:1996}

\begin{equation}
\frac{d\sigma_{NEB}}{dk_1}=\int d^{2}\mathbf{b}_{C}~|S_{CT}\left(\mathbf 
{b}_{C}\right) |^2 \int d^{2}\mathbf{r}_{{Cn}_{\perp}}~\left( 1-|S_{n}\left( \mathbf{b}
_{n}\right )|^2\right)  |{\tilde \phi}_{0}\left(  \mathbf{r}_{{Cn}_{\perp}},k_1\right)|^2.
\label{ek35}%
\end{equation}

\section {Derivation of the STC transfer amplitude }

The derivations of Eq.(\ref{1}) is given in these appendices
in a 
systematic approach which is useful for making 
different approximations to the FSI.
The core spectator model is at the basis of semiclassical methods of breakup and transfer as shown by Eq.(\ref{totx}).
To simplify the notation in this section  the coordinate $\bf r$ and $\bf R$ are used instead of $\bf r_n$ and $\bf r_p$ of Eqs.(\ref{1},\ref{wf1}) and Fig.\ref{fig:IAV_coordinates}.

Semi-classical formulae for transfer amplitudes in heavy ion transfer
reactions which incorporate the kinematical conditions   are developed  following
the approach of Brink \cite{DMB}, Hasan and Brink \cite{hasan,Hasan_1979}, Lo Monaco
and Brink \cite{Monaco_1985} and Bonaccorso et al. \cite{bb,27}. The method of Broglia et
al.\cite{broglia} is very similar to ours. The theory is semi-classical in the sense that
the nuclei involved in the reaction are assumed to follow classical
trajectories but the transfer is calculated by quantum mechanics. The wave
function $\psi$ of the transferred particle satisfies a time-dependent
Schr\"odinger equation 
\begin{equation} {\it i }\hbar {\partial \psi \over \partial t} = (T + V_1(\bra,t) +V_2(\brb,t))\psi. 
\label{(3.18)}\end{equation}
Here $T=-(\hbar^2/2m) \nabla^2$ is the kinetic energy operator for the
transferred particle, while 
the potentials $V_1$ and $V_2$ represent its interaction with
the projectile and target.  The potentials are time dependent and move along
classical trajectories $\bs _1(t)$ and $\bs _2(t)$ describing the motion of
the projectile and target during the collision. 
\begin{equation}V_1({\br}_1,t)=V_1({\br} - \bsa(t)),  \qquad V_2({\br}_2,t) =
V_2({\br} - \bsb(t)).\label{(3.19)}\end{equation}
The initial state $\psi_1(\br,t)$ of the transferred particle satisfies the
time-dependent Schr\"odinger equation for the potential $V_1$ with a 
correction $\Delta V_2$ which takes into account some of the effects of 
$V_2$a
\begin{equation}{\it i} \hbar {\p \psi_1\over \p t} = (T + V_1(\br,t) +\Delta V_2(\br,t))\psi_1.
\label{(3.20)}\end{equation}
In the case of neutron transfer we  can choose $\Delta V_2 =0$, but for
charged particle transfer it is non zero and takes into account the long
range effects of the Coulomb field. The exact form of $\Delta V_2$ will be
specified later. The final state wave-function satisfies a similar equation
\begin{equation}\i \hbar {\p \psi_2\over \p t} = (T + V_2(\br,t) +\Delta V_1(\br,t))\psi_2.
\label{(3.21)}\end{equation}
A very similar starting point as the present one was used in Ref.\cite{YABANA1} where however the eikonal formalism was eventually followed.
During transfer the particle is affected by both the potentials $V_1$ and 
$V_2$ and a perturbation formula \cite{DMB}\cite{broglia} can be
derived for the transfer amplitude $A_{21}$ between the initial state of
the particle $\psi_1$ in the projectile and the final state $\psi_2$ in the
target. The amplitude is 
\begin{equation} A_{21} \approx {1 \over {i \hbar}} \int d t \int d \br
\psi_2^\star(\br ,t) (V_1(\br)- \Delta V_1(\br)) \psi_1 (\br ,t).\label{3.22}\end{equation}

Eq.(\ref{3.22}) can be obtained from the following argument. From Eqs.(\ref{(3.18)}) and  
and (\ref{(3.21)}) we have 
\begin{eqnarray}\i \hbar {\p \over \p t}\langle \psi_2(t)|\psi(t)\rangle=
\langle \psi_2| (T + V_1 +V_2  -(T +V_2 +\Delta V_1))|\psi \rangle =
\langle \psi_2|(V_1 -\Delta V_1) |\psi\rangle.\label{(3.23)}\end{eqnarray}
The state $\psi(t)$ satisfies the initial condition $\psi(t) \rightarrow
\psi_1(t)$ as $t \rightarrow -\infty$. Integrating both sides of Eq.(\ref{(3.23)}) 
gives 
\begin{equation}A_{21} = \lim_{t\rightarrow \infty} \langle \psi_2(t)|\psi(t)\rangle=
{1\over\i \hbar} \int_{-\infty}^{\infty}d t
\langle\psi_2(t)|(V_1 -\Delta V_1) |\psi(t)\rangle.\label{(3.24)}\end{equation}
Eq.(\ref{3.22}) is obtained by making the approximation $\psi(t) \approx 
\psi_1(t)$ in the integral on the right hand side of Eq.(\ref{(3.24)}). An 
alternative derivation of Eq.(\ref{3.22}) is given in the following.

It was shown by Lo Monaco and Brink \cite{Monaco_1985} and Stancu and Brink \cite{ica} that
this perturbation integral can be transformed   to a surface integral over
a surface $\Sigma$ drawn between the two nuclei perpendicular to the line
joining their centers at the point of closest approach. This surface is at
a distance $d_1$ from the center of the first nucleus and $d_2$ from the
second and $d_1 + d_2 = d$. The surface integral formula is 
\begin{equation} A_{21} = {\hbar \over 2m i } \int d t \int d {\bf S} \cdot(\psi_2^\star(\br ,t)\nabla
\psi_1( \br ,t) - \psi_1(\br ,t)\nabla \psi_2^\star( \br ,t)). \label{(3.25)}\end{equation}
Eq.(\ref{(3.25)}) is derived from Eq.(\ref{3.22}) in Appendix C. The two equations are 
exactly equivalent if $V_2 -\Delta V_2=0$ in the region $R_1$ above the 
surface $\Sigma$  and $V_1 -\Delta V_1 =0$ in $R_2$ below 
$\Sigma$. Otherwise Eq.(\ref{(3.25)}) is an approximation to Eq.(\ref{3.22}).

We want to evaluate the amplitude Eq.(\ref{(3.25)}) when the relative motion of the 
two nuclei is a Coulomb orbit. If the scattering angle is small then such 
an orbit can be replaced by a constant velocity orbit tangential to it at 
the point of closest approach. This is a reasonable approximation because 
the transfer takes place near the point of closest approach and because the 
acceleration in the Rutherford orbit is small. It would not be a good 
approximation for large-angle scattering. The transfer amplitude depends 
only on the relative velocity so we can assume that $V_2$ is at rest and 
$V_1$ has a velocity $\bv$ which is the tangential relative velocity in the 
Rutherford orbit at the distance of closest approach $d$. We write the
equation of the orbit relative to the center of $V_2$ as 
	\begin{equation}{\bf R}(t) = {\bf d} + \bv t. \label{(3.26)}\end{equation}
	We will choose a cordinate system so that, at the point of
closest approach between the two nuclei, the z-axis is parallel to the
velocity of the first nucleus relative to the second and the x-axis is in
the reaction plane directed from the center of second nucleus towards the
center of the first. The y-axis is perpendicular to the reaction plane.

\section {Green's function formalism}

We consider the unitary time dependent propagator $G(t_2,t_1)$ which
transforms the solution of the Schr{\"o}dinger equation  \ref{(3.18)} at time $t_1$ into
the solution at time $ t_2$, 
\begin{equation}\psi(t_2)=G(t_2,t_1)\psi(t_1). \label{(A.1)}\end{equation}
 This propagator satisfies the integral equation
\begin{eqnarray}G(t_2,t_1)=G_0(t_2,t_1)+{1\over i\hbar}\int_{t_1}^{t_2}dt ~G_0(t_2,t)
(V_1(t)+V_2(t))G(t,t_1) \label{(A.2)}\end{eqnarray}
which we abbreviate by
$$G=G_0+G_0(V_1+V_2)G.$$
In Eqs.(\ref{(A.2)}) and (\ref{(A.3)}) $G_0$ is the free propagator. Both G and $G_0$ satisfy
the boundary condition
\begin{equation} G(t,t)=G_0(t,t)=1. \label{(A.3)}\end{equation}
First we consider the case of a transition from an initial state 
$\psi(t_1)$ which is a bound state in the potential $V_1$ to a final state
$\psi(t_2)$ which is a bound state in $V_2$. The transition amplitude 
is given by the matrix element 
\begin{equation}A= \lim_{{t_2\to \infty}\atop {t_1\to -\infty}}
\langle\psi_2(t_2)|G(t_2,t_1)|\psi_1(t_1)\rangle\label{(A.4)}\end{equation}
where $\psi_1(t)$ and $\psi_2(t)$ propagate in time according to the 
Schr{\"o}dinger equations (\ref{(3.20)}) and (\ref{(3.21)}) with potentials $V_1 +\Delta 
V_2$ and $V_2 + \Delta V_1$ respectively.
All the multiple scattering effects are included in the propagator
$G$.
The integral equation (\ref{(A.2)}) can be solved by iteration to give a multiple
scattering series. Various partial summations can be made by introducing
the propagators $G_1$ and $G_2$ for the potentials  $V_1 +\Delta V_2$ 
and $V_2+\Delta V_1$ respectively.
\begin{equation}G_1=G_0+G_0(V_1 +\Delta V_2)G_1=G_0+G_0(V_1+\Delta V_2)G_0 +
G_0(V_1+\Delta V_2)G_0(V_1+\Delta V_2)G_0+...\label{ (A.5a)}
\end{equation} 

\begin{equation}G_2=G_0+G_0(V_2+\Delta V_1)G_2 =G_0+G_0(V_2+\Delta V_1)G_0+
G_0(V_2+\Delta V_1)G_0(V_2+\Delta V_1)G_0+... \label{(A.5b)}\end{equation}
The operator $G_1$ is the exact propagator for a nucleon interacting with the
potential $V_1+\Delta V_2$ and similarly for $G_2$. Thus
\begin{equation}\psi_1(t_2) = G_1(t_2,t_1)\psi_1(t_1);~~ 
\psi_2(t_2) = G_2(t_2,t_1)\psi_2(t_1). \label{(A.6)}\end{equation}
The exact propagator G can be written in terms of $ G_2$ or $G_1$ as
\begin{equation}G=G_2+G_2(V_1-\Delta V_1)G=G_2+G(V_1-\Delta V_1)G_2,\label{(A.7a)}\end{equation}
\begin{equation}G=G_1+G_1(V_2-\Delta V_2)G=G_1+G(V_2-\Delta V_2)G_1.\label{(A.7b)}\end{equation}
The transition amplitude equation (\ref{(A.4)}) can be calculated in different
approximations.
Using equations (\ref{(A.6)}) and (\ref{(A.7b)}) we get
\begin{equation}A=\langle\psi_2(t_2)|G_2(t_2,t_1)|\psi_1(t_1)\rangle+\langle\psi_2(t_2)
|G_2(V_1-\Delta V_1)G|\psi_1(t_1)\rangle.
\label{(A.8)}\end{equation}
The first term in equation (\ref{(A.8)}) tends to zero as $t_2 \to \infty$
because of equation  (\ref{(A.6)}). In fact using equation (\ref{(A.6)}) the first term
in equation  (\ref{(A.8)}) can be written as $\langle\psi_2(t_2)|\psi_1(t_2)\rangle.$
This tends to zero as $t_2 \to \infty$ because $\psi_1$ is a bound state in 
the potential $V_1$ which moves away from the target as $t$ increases.
In the second term in equation  (\ref{(A.8)})  $G$ can be written in terms of its
multiple scattering series. Each partial summation represents an
approximation to equation  (\ref{(A.8)}). The simplest 
approximation is obtained by replacing $G$ by $G_1$ 
\begin{equation}A_{21}=\langle\psi_2(t_2)|G_2(V_1-\Delta V_1)G_1|\psi_1(t_1)\rangle,\label{ (A.9)}\end{equation}
 using equation (\ref{(A.6)}) this can be written as
\begin{equation}A_{21}={1\over 2\hbar}\int_{-\infty}^{\infty}~dt~
\langle\psi_2(t)|(V_1-\Delta V_1)|\psi (t)\rangle.\label{(A.10)}\end{equation}
 This is the same as Eq.(\ref{3.22}) above. An equivalent formula can be obtained by 
using Eq. (\ref{(A.7a)}) and approximating $G$ by $G_2$
\begin{equation}A_{21}={1\over 2\hbar}\int_{-\infty}^{\infty}~dt~
\langle\psi_2(t)|(V_2-\Delta V_2)|\psi (t)\rangle.\label{(A.11)}\end{equation}

The argument is slightly different if the final state is a continuum state.
We introduce an asymptotic wave function $\psi_{2out}(t)$ which propagates
in time according to the free particle Schr{\"o}dinger equation (Newton \cite{80}) and satisfies the boundary condition 
$\psi_{2out}(t)\rightarrow \psi_{2}(t)$ as $t\rightarrow \infty$. 
The transition amplitude  (\ref{(A.4)}) can be written in terms of the asymptotic
wave function as
\begin{equation}A=\lim_{{t_2\to \infty}\atop {t_1\to -\infty}}\langle\psi_{2out}(t_2)|G(t_2,t_1)|
\psi_{1i}(t_1)\rangle \label{(A.12)}\end{equation}
It is important to note that $\psi_{2out}(t)$ is a free nucleon wave function.
All the FSI of the nucleon are included in the propagator
$G$.

The transition amplitude equation  (\ref{(A.12)}) can be calculated in different
approximations.
Using equations  (\ref{(A.6)}) and  (\ref{(A.7b)}) we get
\begin{equation}A=\langle\psi_{2out}(t_2)|G_1(t_2,t_1)|\psi_{1i}(t_1)\rangle+\langle\psi_{2out}(t_2)
|GV_2G_1|\psi_{1i}(t_1)\rangle
\label{(A.13)}\end{equation} 
As in the case of Eq. (\ref{(A.8)}) the first term in equation  (\ref{(A.13)})
tends to zero as $t_2 \to \infty$. In the second term in equation  (\ref{(A.13)}) 
$G$ can be written in terms of its multiple scattering series.
Each partial summation represents an approximation to equation  (\ref{(A.13)}). We
consider two cases for the purpose of studying transfer to the continuum. 

i) The propagator $G$ in Eq.(\ref{(A.13)}) is approximated by $G_2$. Then the
transition amplitude  can be written as 
\begin{equation}A_{21}=\langle\psi_{2out}(t_2)|G_2(V_2-\Delta V_2)G_1|\psi_{1i}(t_1)\rangle\label{(A.14)}\end{equation}
when $t_2 \to \infty$. In this matrix element the propagator $G_2$ acting
 on $\psi_{2out}$
introduces FSI with $V_2 +\Delta V_1$. The combination
  \begin{equation}\langle\psi_{2out}(t_2)|G_2(t_2,t)=\langle\psi_{2}^{\prime}(t)| \label{(A.15)}\end{equation}
is a time dependent wave
function for a nucleon propagating in the potential $V_2+\Delta V_1$ of the
target.  Using equations  (\ref{(A.6)}) and  (\ref{(A.15)}),  (\ref{(A.14)}) can be written as 
\begin{equation}A_{21}={1\over 2\hbar}\int_{-\infty}^{\infty}~dt~\langle\psi_{2}^{\prime}(t)|V_2|\psi_{1i}(t
)\rangle.\label{(A.16)}\end{equation}
 This is the case discussed by Bonaccorso and Brink \cite{bb}.

ii. The propagator $G$ in Eq.(\ref{(A.13)}) is approximated by $G_1$. Then the
transition amplitude  can be written as 
\begin{equation}A_{21}=\langle\psi_{2out}(t_2)|G_1(V_2-\Delta V_2)G_1|\psi_{1i}(t_1)\rangle.\label{(A.14bis)}\end{equation}
This case corresponds to inelastic excitations of the projectile.

iii. The propagator $G$ is approximated by the free particle propagator $G_0$,
then one would calculate
\begin{equation}A_{11}=\langle\psi_{2out}(t_2)|G_0V_2G_1|\psi_{1i}(t_1)\rangle.\label{ (A.17)}\end{equation}
Written more explicitly this is
\begin{equation}A_{01}= \int_{-\infty}^{\infty}d t 
\langle \psi_0|(V_2 -\Delta V_2)|\psi_1 \rangle.\label{(A.17)}\end{equation}
This form corresponds to the break-up of
the nucleon in the projectile due to the interaction with the target.

Case i) includes the FSI of the nucleon with the potential
$V_2$. It can be used to study the influence of nucleon resonances in the target
nucleus.  Similarly case ii) includes the final state interaction of the neutron with the potential of the projectile and can be used to study effect of resonances in the projectile. There are no FSI in approximation iii).

Eq.( \ref{(A.10)}) can be transformed into a simpler form in the case of a
peripheral collision. Let $\Sigma$ be a surface which lies between the two
potentials $V_1$ and $V_2$ and which divides the space into regions $R_1$ and
$R_2$. In the following we assume that 
\begin{equation}V_1 -\Delta V_1 =0 \quad {\rm in}~R_2;\qquad V_2 -\Delta V_2 =0
\quad{\rm in}~R_1.\label{(A.19)}\end{equation}
Then the matrix element 
$\langle \psi_2 |(V_1 -\Delta V_1)|\psi_1 \rangle$ in eq  (\ref{(A.10)}) can be 
written as a sum of integrals over the regions $R_1$ and $R_2$. The 
integral over $R_2$ is zero if $V_1 -\Delta V_1 =0$ in $R_2$. Using 
Eq.(\ref{(3.20)}) the first integral can be written as 
\begin{eqnarray}\int_{R_1} \psi_2^* (V_1 - \Delta V_1)\psi_1 d^3 \br
= \int_{R_1} \psi_2^* \Bigl(\i \hbar{\p \over \p t} 
+{\hbar^2 \over 2m} \nabla^2  -\Delta V_2 - \Delta V_1\Bigr)\psi_1 d^3 \br.\end{eqnarray}
Applying Green's theorem reduces this to 
\begin{eqnarray}\int_{R_1} \psi_2^* (V_1 - \Delta V_1)\psi_1 d^3 \br
=&&{\hbar^2 \over 2m}\int_{\Sigma}d {\bf S}\cdot
(\psi_2^*\nabla \psi_1 - \psi_1 \nabla \psi_2^*)+ \i \hbar {\p \over \p t}\int_{R_1} \psi_2^* \psi_1 d^3 \br \nonumber  \\
&+&  \int_{R_1} \Bigl(\i \hbar{\p \psi_2 \over \p t}+{\hbar^2 \over 2m} \nabla^2 \psi_2  
-\Delta V_2 \psi_2 - \Delta V_1 \psi_2\Bigr)^*\psi_1 d^3 \br.\label{(AA.19)}\end{eqnarray}
where $d {\bf S}$ is a surface element normal to $\Sigma$ directed out of
$R_1$. If Eq.(\ref{(AA.19)}) is integrated over time between $t=-\infty$ and 
$t=\infty$ the second term vanishes because the potentials $V_1$ and $V_2$ 
are very far away from each other as $t \to \pm \infty$ and $\psi_1$ and 
$\psi_2$ have no overlap in that limit. If $V_2 -\Delta V_2=0$ in $R_1$ 
then the third term in Eq.(\ref{(AA.19)}) is zero because $\psi_2$ satisfies the 
Schr{\"o}dinger equation  (\ref{(3.21)}). This proves Eq.(\ref{(3.25)}).
The next step is to obtain our final formula Eq.(\ref{(3.30)}) from (\ref{(3.25)}). 

First the wave functions 
$\psi_1$ and $\psi_2$ are expressed in terms of $\phi_1$ and $\phi_2$ using 
Eq.(\ref{wf1}). They can be written as inverse transforms of momentum distributions
\begin{equation}\phi_1(x,y,z) ={1\over (2\pi)^2}\int\int d k_{1y} d k_{1z}
\exp^{\i( y k_{1y} +z k_{1z})}\tilde{\phi}(x,k_{1y},k_{1z}),\label{(A.20)}\end{equation}
and similarly for $\phi_2(x,y,z)$. Substituting into Eq.(\ref{(3.25)}) yields a 
7-dimensional integral over $y,~z,~t$ and $k_{1y},~k_{1z},~k_{2y},~k_{2z}$. 
This integral contains a phase factor 
\begin{equation}\exp [mvz/\hbar + (\varepsilon_2 -\varepsilon_1 -\half m v^2)t /\hbar
+(z-vt)k_{1z} +y k_{1y}-zk_{2z} -yk_{2y}] \end{equation}
which contains all the dependence on $y$, $z$ and $t$. Hence the integration
over those variables gives a product of three $\delta$-functions 
\begin{equation}{(2\pi)^3 \over v}\delta (k_{1z} -k_{2z} +mv/\hbar)
\delta (k_{1z} - (\varepsilon_2 -\varepsilon_1 -\half m v^2)/\hbar v)
\delta (k_{1y} - k_{2y}).\label{(A.21)}\end{equation}
Finally Eq.(\ref{(3.25)}) reduces to a 1-dimensional integral over $k_y = k_{1y} = 
k_{2y}$. The remaining two $\delta$-functions in (\ref{(A.21)}) give
\begin{equation}k_{1z} = k_1 = (\varepsilon_2-\varepsilon_1 -\half m v^2)/\hbar v,\\
k_{2z} = k_2 = (\varepsilon_2-\varepsilon_1 +\half m v^2)/\hbar v.\label{(A.22)}\end{equation}
These are the values of $k_1$ and $k_2$ used in the text following Eq.(\ref{1}) and in Eq.(\ref{k12}). After evaluation the derivatives 
normal to the surface $\Sigma$ in \ref{(3.25)}) one obtains
a factor $\i(\gamma_{1x} +\gamma_{2x}) = 2\i 
\sqrt{\eta^2 + k_z^2}$. When these results are collected the final form of the transfer (breakup)  amplitude is obtained as
\begin{equation}A_{21} = {\hbar \over 2\pi mv}\int d k_y \sqrt{\eta^2 +k_y^2}
\tilde{\phi}_2^\star(d_2, k_y, k_2)  \tilde{\phi}_1(-d_1, k_y, k_1).
\label{(3.30)}\end{equation}

Note that the amplitude $A_{21}$ of this section was indicated as $A_{fi}$ in the main text.

%

%

%


%
\end{appendix}


\end{document}